\documentclass[aps,amssymb,10pt,showpacs,showkeys,letterpaper]{revtex4}
\usepackage[utf8]{inputenc}
\usepackage{graphicx}
\usepackage{amsmath}
\usepackage{epsfig}
\usepackage{bm}

\newcommand{\be}{\begin{equation}}
\newcommand{\ee}{\end{equation}}
\newcommand{\beq}{\begin{eqnarray}}
\newcommand{\eeq}{\end{eqnarray}}
\providecommand{\abs}[1]{\lvert#1\rvert}
\begin{document}

\title{Collision of particles near a three-dimensional rotating Ho\v{r}ava AdS black hole}
\author{Ram\'{o}n B\'{e}car}
\email{rbecar@uct.cl}
\affiliation{Departamento de Ciencias Matem\'{a}ticas y F\'{\i}sicas, Universidad Cat\'{o}lica de Temuco, Montt 56, Casilla 15-D, Temuco, Chile}
\author{P. A. Gonz\'{a}lez}
\email{pablo.gonzalez@udp.cl}
\affiliation{Facultad de Ingenier\'{\i}a y Ciencias, Universidad Diego Portales, Avenida Ej\'{e}rcito Libertador 441, Casilla 298-V, Santiago, Chile.}
\author{Yerko V\'{a}squez}
\email{yvasquez@userena.cl}
\affiliation{Departamento de F\'{\i}sica, Facultad de Ciencias, Universidad de La Serena,\\
Avenida Cisternas 1200, La Serena, Chile.}
\date{\today}

\begin{abstract}

We consider a three-dimensional rotating Ho\v{r}ava AdS black hole, that corresponds to a Lorentz-violating version of the BTZ black hole and we analyze the effect of the breaking of Lorentz invariance on the possibility that the black hole can acts as a particle accelerator by analyzing 
the energy in the center of mass (CM) frame of two colliding particles in the vicinity of its horizons. We find that the critical angular momentum of particles increases when the Ho\v{r}ava parameter $\xi$ increases and when the aether parameter $b$ increases. Also, the particles can collide on the inner horizon with arbitrarily high CM energy if one of the particles has a critical angular momentum being possible the BSW process, for the non-extremal rotating Ho\v{r}ava AdS black hole. Mainly, while that for the extremal BTZ black hole the particles with critical angular momentum only can exist on the degenerate horizon, for the Lorentz-violating version of the BTZ black hole the particle with critical angular momentum can exist in a region from the degenerate horizon.

\end{abstract}

\maketitle


\newpage

\tableofcontents


\newpage

\section{Introduction}

Ba\~nados, Silk and West (BSW) \cite{Banados:2009pr}  demonstrated that two particles colliding near the degenerate horizon of an extreme Kerr black hole could create a large center of mass (CM) energy if one of the particles has a critical angular momentum; thus, extreme Kerr black holes can act as natural particle accelerators. Nowadays, this process is known as the BSW mechanism, which was found for the first time by Piran, Shaham and Katz in 1975 \cite{PS1,Piran:1977dm,PS3}. The same mechanism also occurs
for non-extremal ones \cite{Grib:2010at}, and
it is a universal property of rotating black holes 
\cite{Zaslavskii:2010jd}. Also, it was shown that the similar effect exists for non-rotating charged black holes 
\cite{Zaslavskii:2010aw}. Moreover, the extension of the BSW mechanism to non-extremal backgrounds shows that particles cannot collide with arbitrarily high energies at the outer horizon and that ultra-energetic collisions  can only occur near the Cauchy horizon of a Kerr black hole with any spin parameter \cite{Gao:2011sv}.
The non-extremal Kerr-de Sitter black holes could also act as particle accelerators with arbitrarily high CM energy if one of the colliding particles has the critical angular momentum \cite{Li:2010ej}. The BSW mechanism has been extensively studied for different black hole geometries  \cite{Said:2011qm, Zhong:2011vq, Yi:2011,  Abdujabbarov:2013qka,  Galajinsky:2013as,  Shaymatov:2013tna, Sadeghi:2013gmf, Fernando:2013qba, Ghosh:2014mea, Pradhan:2014eza,
Ghosh:2015pra, Amir:2015pja, Guo:2016vbt, Zhang:2016btg, Armaza:2015eha, Zaslavskii:2016stw, Zaslavskii:2016dfh, Fernando:2017kut, Gonzalez:2018lfs, Gonzalez:2018zuu, Jiang:2019cuc}.
Furthermore, the formation of black holes through the BSW mechanism was investigated in \cite{ata}. \\

Nowadays, one could think that Lorentz invariance may not be fundamental or exact, but is merely an emergent symmetry on sufficiently large distances or low energies. It has been suggested in Ref. \cite{Horava:2009uw} that giving up Lorentz invariance by introducing a preferred foliation and terms that contain higher-order spatial derivatives can lead to significantly improved UV behavior, the corresponding gravity theory is dubbed Hořava gravity. The three-dimensional Hořava gravity \cite{Sotiriou:2011dr} admits a Lorentz-violating version of the BTZ black hole, i.e. a black hole solution with AdS asymptotics, only in the sector of the theory in which the scalar degree of freedom propagates infinitely fast \cite{Sotiriou:2014gna}. Remarkably, in contrast to general relativity, the three-dimensional Hořava gravity also admits black holes with positive and vanishing cosmological constant. The propagation of a massive scalar field is stable in this background  \cite{Becar:2019hwk}. Also, new kinds of orbits are allowed, such as unstable circular orbits and trajectories of first kind for the motion of photons \cite{Gonzalez:2019xfr}. The aim of this work is to consider three-dimensional rotating Ho\v{r}ava AdS black holes \cite{Sotiriou:2014gna} and to study, via the BSW mechanism, the possibility of obtaining unbounded energy in the CM frame of two colliding particles 
and to analyze the effect of Lorentz breaking symmetry on this. The BSW mechanism in Ho\v{r}ava-Lifshitz black hole was studied in four spacetime dimensions \cite{Abdujabbarov:2011af, Sadeghi:2011qu}, and in two spacetime dimensions  \cite{Halilsoy:2015rna}. 
It is worth mentioning that for extreme rotating black holes in four-dimensional Ho\v{r}ava gravity the fundamental parameter of Ho\v{r}ava gravity can avoids an infinite value of the CM energy  \cite{Abdujabbarov:2011af}.\\

In three spacetime dimensions the collision of two particles near the horizon of a BTZ black hole was studied in Ref. \cite{Lake:2010bq, Yang:2012we, Tsukamoto:2017rrl}. In Ref. \cite{Lake:2010bq,Yang:2012we} the authors found that the particle with the critical angular momentum could exist inside the outer horizon of the BTZ black hole regardless of the particle energy with the BSW process being possible on the inner horizon for the non-extremal BTZ black hole. Also, the BSW process could also happen for the extremal BTZ black hole, where the particle with the critical angular momentum could only exist on the degenerate horizon. On the other hand, in Ref. \cite{Tsukamoto:2017rrl}, the authors studied the collision of two particles on the event horizon and outside of the BTZ black hole, and they showed that although in principle the CM energy of two ingoing particles can be arbitrarily large on the event horizon, if either of the two particles has a critical angular momentum and the other has a non-critical angular momentum, the critical particles never reach the event horizon. However, the motion of a particle with a subcritical angular momentum is allowed near an extremal rotating BTZ black hole and that a CM energy for a tail-on collision at a point can be arbitrarily large in a critical angular momentum limit. Also, the BSW effect is possible on the outer horizon in the extremal warped AdS$_3$ black hole, and the particle with critical angular momentum can reach the degenerate horizon when a condition on its energy is fulfilled \cite{Becar:2017aag}, which resembles to what occurs in the extremal Kerr-AdS black hole; however, in the extremal Kerr-AdS black hole two conditions must be fulfilled \cite{Li:2010ej}; besides, this effect is also possible on the inner horizon for the non-extremal warped AdS$_3$ black hole. \\

The manuscript is organized as follows: In Sec. \ref{Background} we give a brief review of the three-dimensional rotating Ho\v{r}ava black holes. Then, we study the particle's motion in this background in Sec. \ref{motion}. In Sec. \ref{CMS}, we obtain the CM energy of two colliding particles, and in Sec. \ref{two} we study the radial motion of a particle with critical angular momentum and we investigate the possibility that
the black hole acts as a particle accelerator.  Finally, our conclusions are in Sec. \ref{conclusion}. \\

\section{Three-dimensional rotating Ho\v{r}ava black holes}
\label{Background}

\begin{figure}[!h]
\begin{center}
\includegraphics[width=100mm]{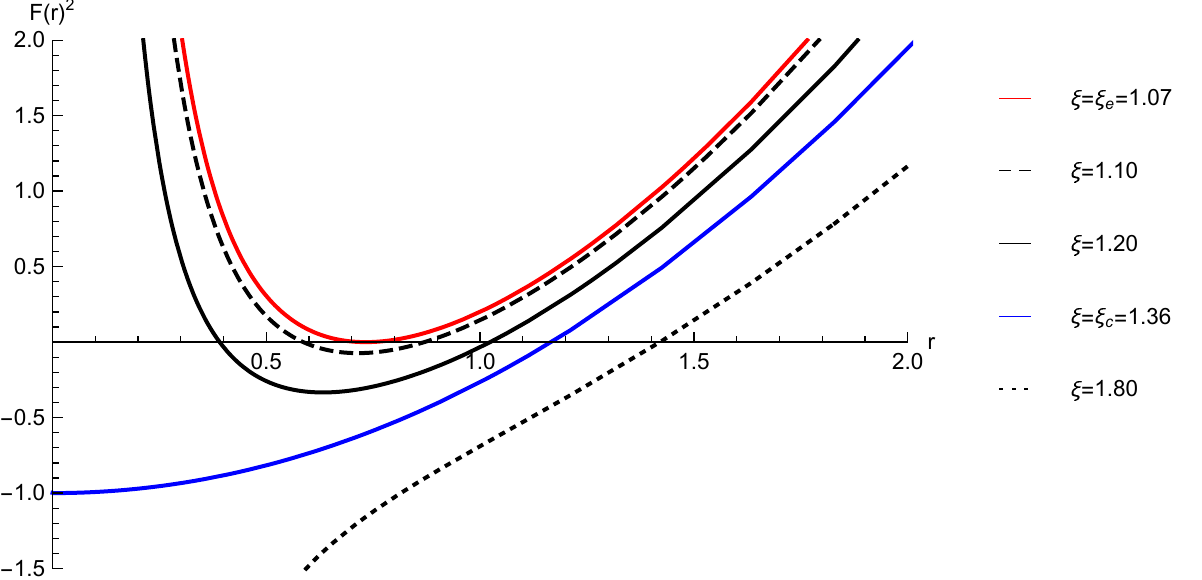}
\end{center}
\caption{The behavior of the lapse function as a function of $r$, with $M=1$, $\lambda=1$, $a=1$, $b=1$, $\Lambda=-1$, $J=1.2$, for different values of $\xi$. }
\label{fl}
\end{figure}

The three-dimensional Ho\v{r}ava gravity is described in a preferred foliation by the action  \cite{Sotiriou:2011dr}
\begin{equation}
S_{H}=\frac{1}{16\pi G_{H}}\int dT d^2x N\sqrt{g}\left[L_{2}+L_{4}\right] \,,
\end{equation}
being the line element in the preferred foliation
\begin{equation}
    ds^2=N^2dT^2-g_{ij}(dx^i+N^idT)(dx^j+N^jdT)\,,
\end{equation}
where $g_{ij}$ is the induced metric  on the constant-$T$ hypersurfaces. $G_{H}$ is a coupling constant with dimensions of a length squared, $g$ is the determinant of $g_{ij}$ and the Lagrangian $L_{2}$ has the following form
\begin{equation}
 L_{2}=K_{ij} K^{ij}-\lambda K^2+\xi\left(^{(2)}R-2\Lambda\right)+\eta a_{i} a^{i}\,,
\end{equation}
where $K_{ij}$, $K$, and $^{(2)}R$ correspond to extrinsic, mean, and scalar curvature, respectively, and $a_{i}$ is a parameter related to the lapse function $N$ via $a_{i}=-\partial_{i}\ln{N}$. $L_4$ corresponds to the set of all the terms with four spatial derivatives that are invariant under diffeomorphisms. For $\lambda=\xi=1$ and $\eta=0$,  the action reduces to that of general relativity. In the infrared limit of the theory the higher order terms $L_{4}$ (UV regime) can be neglected, and the theory is equivalent to a restricted version of the Einstein-aether theory, the equivalence can be showed by restricting the aether to be hypersurface-orthogonal and the following relation is obtained
\begin{equation}
    u_\alpha=\frac{\partial_\alpha T}{\sqrt{g^{\mu\nu}\partial_\mu T \partial \nu T}}\,,
\end{equation}
where $u_\alpha$ is a unit-norm vector field called the aether, see Ref. \cite{Jacobson:2010mx} for details. Another important characteristic of this theory is that only in the sector $\eta=0$, Ho\v{r}ava gravity admits asymptotically AdS solutions \cite{Sotiriou:2014gna}. Therefore, assuming stationary and circular symmetry the most general metric is given by
\begin{equation}\label{metric2}
ds^{2}=Z(r)^2dt^{2}-\frac{1}{F(r)^2}dr^{2}-r^{2}(d\phi+\Omega(r)dt)^2~,
\end{equation}
and by assuming the aether to be hypersurface-orthogonal, it results
\begin{equation}
    u_t=\pm \sqrt{Z(r)^2(1+F^2(r)U^2(r))}\,,u_r=U(r)\,.
\end{equation}
The theory admits the BTZ ``analogue'' to the three-dimensional rotating Ho\v{r}ava black holes described by the solution
\begin{equation}
F(r)^2= Z(r)^2 =-M +\frac{\bar{J}^2}{4r^2}-\bar{\Lambda}r^2~,~
\Omega(r)=-\frac{J}{2 r^2} \,,~
    U(r) =\frac{1}{F(r)}\left(\frac{a}{r}+br\right)\,,
\end{equation}
where
\begin{equation}
\bar{J}^2=\frac{J^2+4a^2(1-\xi)}{\xi}~,~\bar{\Lambda}=\Lambda-\frac{b^2(2\lambda -\xi-1)}{\xi}~.
\end{equation}
The sign of the effective cosmological constant $\bar{\Lambda}$ determines the asymptotic behavior (flat, dS, or AdS) of the metric. Also, $\bar{J}^2$ can be negative, this occurs when either $\xi < 0$ or $\xi > 1$, $a^2 > J^2/(4(\xi-1))$.
The aether configuration for this metric is given explicitly by
\begin{equation}
u_t = \sqrt{F^2+\left(\frac{a}{r}+br\right)^2}\,,~ 
u_r = \frac{1}{F^2}\left(\frac{a}{r}+br\right)\,,~ 
u_{\phi} = 0 \,,
\end{equation}
where $a$ and $b$ are constants that can be regarded as measures of aether misalignment, with $b$ as a measure of asymptotically misalignment, such that for $b\neq 0$ the aether does not align with the timelike Killing vector asymptotically. Note that for $\xi=1$ and $\lambda=1$, the solution becomes the BTZ black hole, and for $\xi=1$ and $\lambda \neq 1$, the solution becomes the BTZ black hole with a shifted cosmological constant $\bar{\Lambda}=\Lambda- 2b^2(\lambda-1)$. However, there is still a preferred direction represented by the aether vector
field which breaks Lorentz invariance for $\lambda \neq 1$ and $b\neq 0$. The locations of the inner and outer horizons $r = r_\pm$, are given by
\begin{equation}\label{horizon}\
r_{\pm}^2=-\frac{M}{2\bar{\Lambda}}\left(1\pm \sqrt{1+\frac{\bar{J}^2\bar{\Lambda}}{M^2}} \right)~,
\end{equation}
and the Hawking temperature $T_H$ is given by $T_H=\frac{-\bar{\Lambda}(r_+^2-r_-^2)}{2\pi r_+}$.
For $\bar{J}\ne J$ ($\xi\ne 1$), there is a curvature singularity due to the Ricci scalar $R=-6\bar{\Lambda}+\frac{1}{2 r^2}\left(\bar{J}^2-J^2\right)$ is divergent at $r=0$, in contrast to the BTZ black holes where the Ricci and Kretschmann scalars are finite and smooth at $r=0$.
Considering $M>0$ and a negative cosmological constant $\bar{\Lambda}<0$, the condition $\bar{J}^2 \abs{\bar{\Lambda}}\leq M^2$ must be fulfilled for the solution represents a black hole. For $0<\bar{J}^2\abs{\bar{\Lambda}} < M^2$ the black holes  have inner and outer horizons $r_-$ and $r_+$, the extremal case corresponds to $\bar{J}^2 \abs{\bar{\Lambda}}= M^2$, while that for $\bar{J}^2< 0$ the black holes have outer horizon $r_+$, but no inner horizon $r_-$.
The value of $\xi$ for which the black hole is extremal is given by \cite{Becar:2019hwk}
\begin{eqnarray}
\notag  \xi_e &=&   -\frac{1}{2(M^2-4a^2 (b^2+ \Lambda))} \Big( b^2 (J^2+8a^2 \lambda)+ \Lambda (J^2+ 4 a^2) -\Big( (b^2 (J^2+8a^2 \lambda)+ \Lambda (J^2+ 4 a^2))^2   \\
 &&  +4b^2 (J^2+4 a^2)(2 \lambda-1) (M^2-4 a^2 (b^2 + \Lambda) )\Big)^{1/2}  \Big) \, ,
\end{eqnarray}
the value of $\xi$ for which the black hole turns from having two horizons to having one horizon is given by  $\xi_c = \frac{4 a^2 +J^2}{4 a^2}$,
and the value of $\xi$ for which the effective cosmological constant $\bar{\Lambda}$ changes sign is $\xi=\frac{(2 \lambda-1) b^2}{b^2 + \lambda}$. In Fig. (\ref{fl}), we show the behavior of the lapse function $F(r)^2$, for a choice of parameters, and different values of $\xi$, where we can observe the existence of one horizon for $\xi>\xi_c$ ($\xi=1.80$) with $\bar{J}^2<0$, one horizon with $r_-=0$, for $\xi=\xi_c\approx 1.36$ or $\bar{J}=0$, two horizons for $\xi_e<\xi<\xi_c$, and one degenerate horizon for $\xi=\xi_e\approx 1.07$.
\\
\\

\begin{figure}[!h]
\begin{center}
\includegraphics[width=110mm]{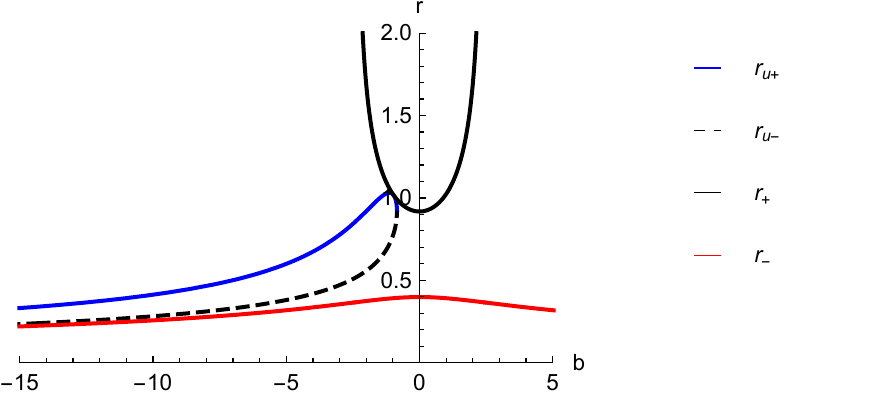}
\end{center}
\caption{The behavior of the horizons as a function of the parameter $b$,  with $M=1$, $\xi=1.2$, $\lambda=1$, $a=1$, $\Lambda=-1$, and $J=1.2$. For $b\approx -0.84$, $r_u^+=r_u^-$.}
\label{Hb}
\end{figure}

\begin{figure}[!h]
\begin{center}
\includegraphics[width=110mm]{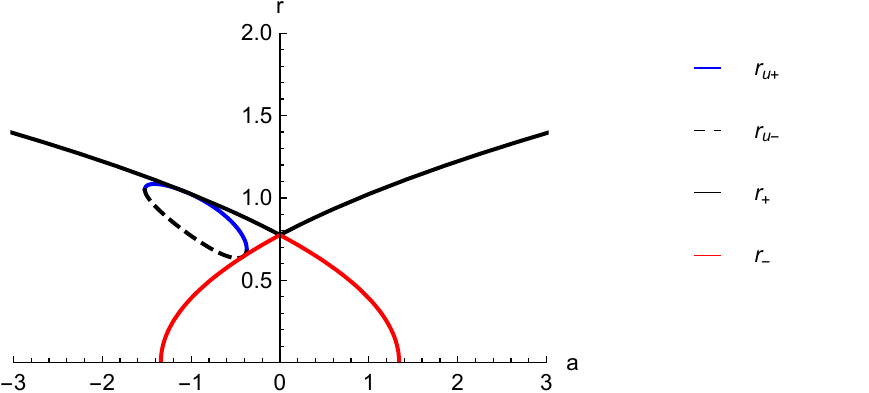}
\end{center}
\caption{The behavior of the horizons as a function of the parameter  $a$,  with $M=1$, $\xi=1.2$, $\lambda=1$, $b=1$, $\Lambda=-1$, and $J=1.2$. For $a\approx -1.52$, and $-0.37$, $r_u^+=r_u^-$.}
\label{Ha}
\end{figure}

Besides the existence of inner and outer horizons, what is perhaps the most interesting feature within the
context of Lorentz-violating gravity theories is that they
can have universal horizons, which are given by \cite{Sotiriou:2014gna}
\begin{equation}\label{universal}\
 (r_{u}^\pm)^2 =\frac{M-2ab}{2(b^2-\bar{\Lambda})}\pm\frac{1}{2(b^2-\lambda)}\left[(M-2ab)^2-(4a^2+\bar{J}^2) (b^2-\bar{\Lambda})\right]^{\frac{1}{2} }\,.
\end{equation}
In Fig. (\ref{Hb}), we plot the behavior of the horizons as a function of the parameter $b$, and as a function of $a$ in Fig. (\ref{Ha}),  for a choice of parameters. It is possible to observe different zones. One of them, is limited by $r_-$ and $r_+$; and it is described by the existence of the aether, where the roots $r_u^{\pm}$ are imaginary and therefore there are no universal horizons. Other zones are characterized by two real and distinct universal horizons inside the region between $r_-$ and $r_+$, outside $r_-$, and inside $r_+$; and an especial zone where both universal horizons coincide and is given by 
\begin{equation}
    r_u^2=\frac{M-2a_{\pm}(M,\bar{J},b)b}{2(b^2-\bar{\Lambda}(b))}\,,
\end{equation}
where $a_{\pm}$ are the roots of
\begin{equation}
    \frac{(4a^2+\bar{J}^2)(b^2-\bar{\Lambda}(b))}{\xi(M-2ab)^2}=1\,.
\end{equation}
In the region between $r_u^-$ and $r_u^+$, the aether turns imaginary and the foliation cannot be extended until the singularity. So, if this region is located between the inner and outer horizon then the BSW process it is not possible on the inner horizon $r_-$ because the aether is imaginary.


\newpage

\section{Motion equations} 
\label{motion}
\begin{figure}[!h]
\begin{center}
\includegraphics[width=80mm]{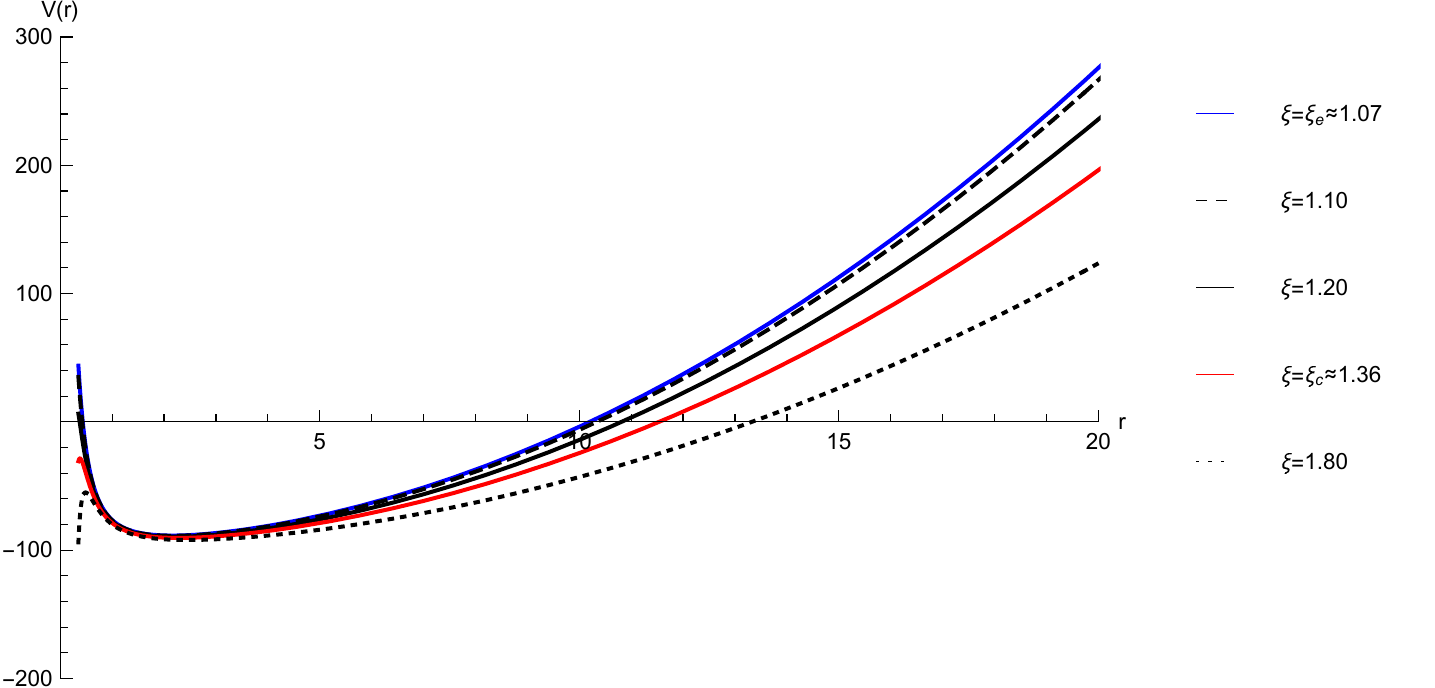}
\includegraphics[width=80mm]{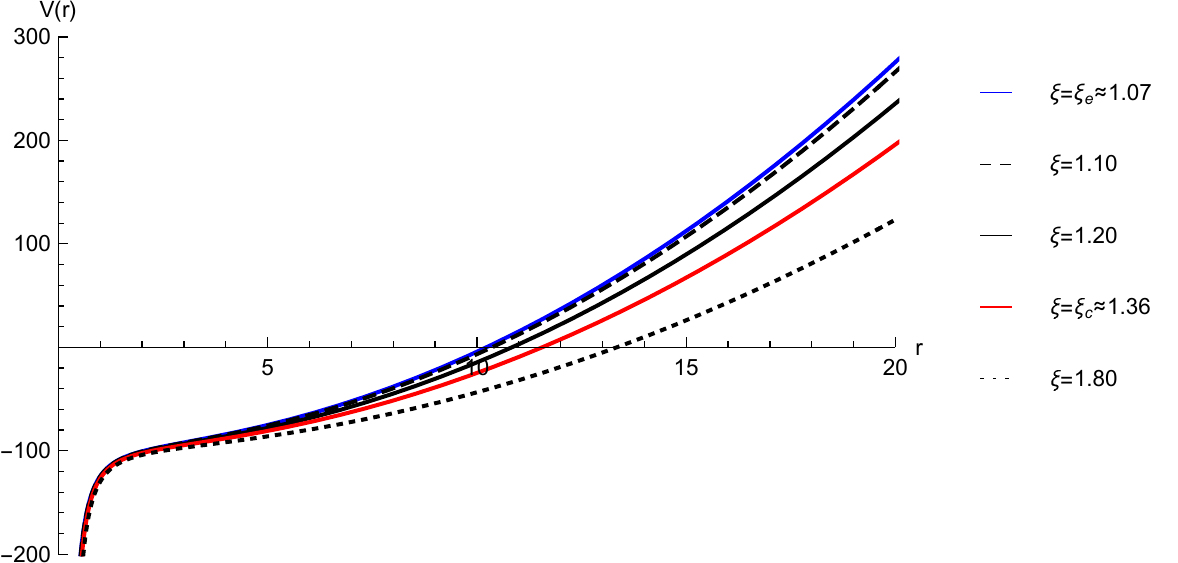}
\end{center}
\caption{The behavior of the effective potential $V(r)$ for massive particles $m=1$ as a function of $r$  for different values of $\xi$ and $M=1$, $\lambda=a=b=1$, $\Lambda=-1$, $J=1.2$, $E=10$. Left figure shows the behavior of the effective potential for $L=2$, and $r>0.35$, where inequality (\ref{condition1}) is satisfied, while the right figure shows the behavior of the effective potential for $L=-2$, where inequality (\ref{condition1}) is satisfied for all $r$.}
\label{f2}
\end{figure}

\begin{figure}[!h]
\begin{center}
\includegraphics[width=80mm]{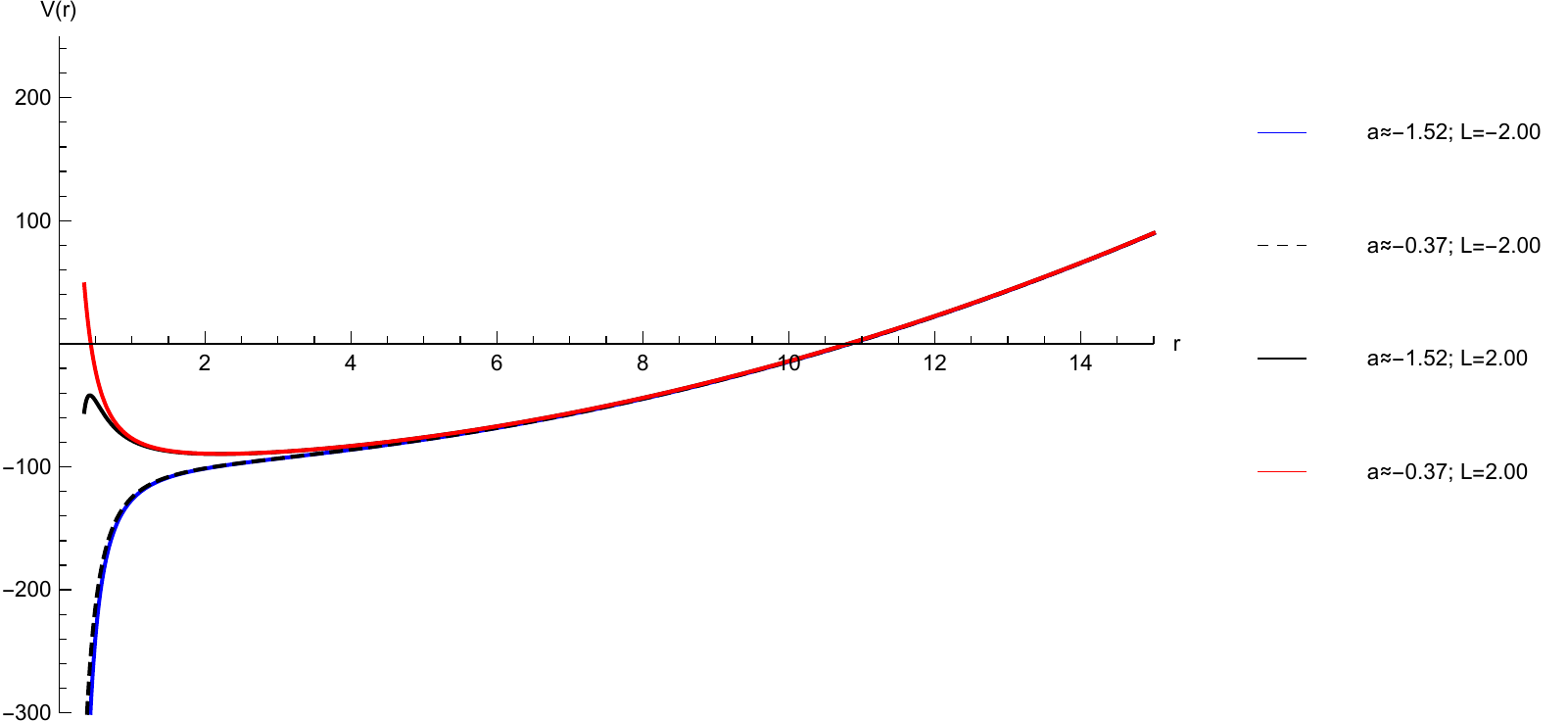}
\end{center}
\caption{The behavior of the effective potential $V(r)$ when $r_u^-=r_u^+$ for massive particles $m=1$ as a function of $r$  with $\xi=1.2$, $M=1$, $\lambda=b=1$, $\Lambda=-1$, $J=1.2$, and $E=10$. For $L=2$,  $r>0.35$, in order to satisfy the inequality (\ref{condition1}), while that for $L=-2$, the inequality (\ref{condition1}) is satisfied for all $r$.}
\label{f22}
\end{figure}

In this section, we find the motion equations of probe massive particles around the three-dimensional Ho\v{r}ava AdS black hole. It is important to emphasize \cite{Cropp:2013sea} that in a Lorentz-violating scenario, particles will be generically coupled to the aether field generating UV modifications of the matter dispersion relations, it is more, one can also expect radiative corrections in the infrared sector but these contributions are suppressed by knows mechanisms. In our analysis we are interested on the infrared limit of the theory; so the presence of higher-order terms ($L_4$) related to the UV behavior of the theory are ignored and in this case the theory can be formulated in a covariant fashion, and it then becomes equivalent to a restricted version of Einstein-aether theory \cite{Sotiriou:2014gna}. Since our analysis is focused in the low energy part of the theory, the interaction between the massive particle and the aether field is ignored, thus the presence of aether field only affect the background spacetime geometry. It is worth mentioning  that a similar analysis was performed in \cite{Zhu:2019ura} where the authors analyzed the evolution of the photon around the static neutral and charged aether black holes using the Hamilton-Jacobi equation. Therefore, the massive particles follow the typical geodesics in such given black holes spacetime, that can be derived from the Lagrangian of a test particle, which is given by \cite{chandra}
\begin{equation}
\mathcal{L}=\frac{1}{2}g_{\mu\nu}\dot{x}^{\mu}\dot{x}^{\nu}~,
\end{equation}
where $\dot{x}^{\mu}=dx^{\mu}/d\tau$, and $\tau$ is an affine parameter along the geodesic that we choose as the proper time. So, for the three-dimensional rotating Ho\v{r}ava AdS black hole described by the metric (\ref{metric2}), the Lagrangian reads
\begin{equation}\label{tl4}
  2\mathcal{L}=\left(Z^2(r)-r^2 \Omega^2(r)\right)\dot{t}^2-\frac{1}{Z^2(r)}\dot{r}-r^2\dot{\phi}^2-2r^2\Omega(r)\dot{t}\dot{\phi}~.
\end{equation}
Since this Lagrangian is
independent of the cyclic coordinates ($t,\phi$), their
conjugate momenta ($\Pi_t, \Pi_{\phi}$) are conserved. The equations of motion are obtained from
$ \dot{\Pi}_{q} - \frac{\partial \mathcal{L}}{\partial q} = 0$, where $\Pi_{q} = \partial \mathcal{L}/\partial \dot{q}$
are the conjugate momenta to the coordinate $q$, and are given by
\begin{equation}
\Pi_{t} = \left(Z^2(r)-r^2 \Omega^2(r)\right)\dot{t}+ -r^2\Omega(r)\dot{\phi}\equiv E~,~
\Pi_{r}= \frac{1}{Z^2(r)}\dot{r}~,~
 \Pi_{\phi}
=  r^2\dot{\phi}-r^2\Omega(r)\dot{t} \equiv -L~,
\label{w.11c}
\end{equation}
where $E$ and $L$ are dimensionless integration constants associated with each of them.
The Hamiltonian
\begin{equation}
\mathcal{H}=\Pi_{t} \dot{t} + \Pi_{\phi}\dot{\phi}+\Pi_{r}\dot{r}
-\mathcal{L}~,
\end{equation}
yields
\begin{equation}
2\mathcal{H}=-E\, \dot{t} + L\,\dot{\phi}-\frac{\dot{r}^2}{Z^2(r)}\equiv m^2~.
\label{w.11z}
\end{equation}
Now, we solve the above equations for $\dot{r}^2$ in order to obtain the radial equation which allows us to
characterize the possible movements of the test particles without an explicit solution of the equations of motion, which yield
\begin{eqnarray}
&&\dot{t}= \frac{E+L\Omega(r)}{Z^2(r)}~,\\
\label{w.12}
&&\dot{\phi}= -\frac{ E \Omega(r)}{Z^2(r)}+L\left( \frac{-\Omega(r)^2}{Z^2(r)}+\frac{1}{r^2} \right)~,\\
\label{w.13}
&&\dot{r}^{2}= Z^2(r) \left(\frac{1}{Z^2(r)} (E+L \Omega(r))^2-\frac{L^2}{r^2}-m^2\right)~.
\label{rdot}
\end{eqnarray}
The above equations represent all nonzero {\bf{3}}-velocity components $u=(\dot{t},\dot{r},\dot{\phi})$ for the geodesic motion that will be used in the next section to obtain the CM energy of two colliding particles falling freely from rest with the same rest mass $m_{0}$ in the three-dimensional rotating Ho\v{r}ava AdS black hole background.
We will assume $\dot{t}>0$ for all $r>r_+$ so that the motion is forward in time outside the horizon. So, the following condition must be fulfilled 
\begin{equation}
\label{condition1}
E+L\Omega(r)>0, \,\, \text{for all}\,\, r>r_+~.
\end{equation}
Now, we write the equation of motion of the particle in the radial direction as:
$\dot{r}^{2}+V(r)=0$, 
where $V$ is the effective potential of the particle in the radial direction and reads
\begin{equation}\label{potencial}
V(r)=-\left( \left( E+L\Omega(r)\right) ^{2}-Z^{2}(r)\left(\frac{L^{2}}{r^2}+m^2\right)\right)~.
\end{equation}
By analyzing this effective potential we can determine if a particle can reach the event horizon. The motion of the particle is allowed in regions where $V(r)\leq 0$, and it is prohibited in regions where $V(r)>0$. It is clear that the particle can exist on the event horizon $r=r_{+}$ because $Z^{2}(r_{+})=0$, and then the effective potential is negative.
On the other hand, when $r \rightarrow \infty$ it is easy to show that the effective potential approximates to
\begin{equation}
V(r \rightarrow \infty) \approx -m^2\bar{\Lambda}r^2~. \label{asym}
\end{equation}
This expression shows that the existence of a massive particle at infinity depends on the effective cosmological constant ($ \bar{\Lambda}$)  and not on its energy $E$. Therefore, massive particles can not exist at infinity in the $AdS$  case, as is shown in Fig. (\ref{f2})  for different values of the parameter $\xi$, and in Fig. (\ref{f22}) for the case $r_u^-=r_u^+$.

\newpage

\section{The CM energy of two colliding particles}
\label{CMS}
\begin{figure}[!h]
\begin{center}
\includegraphics[width=100mm]{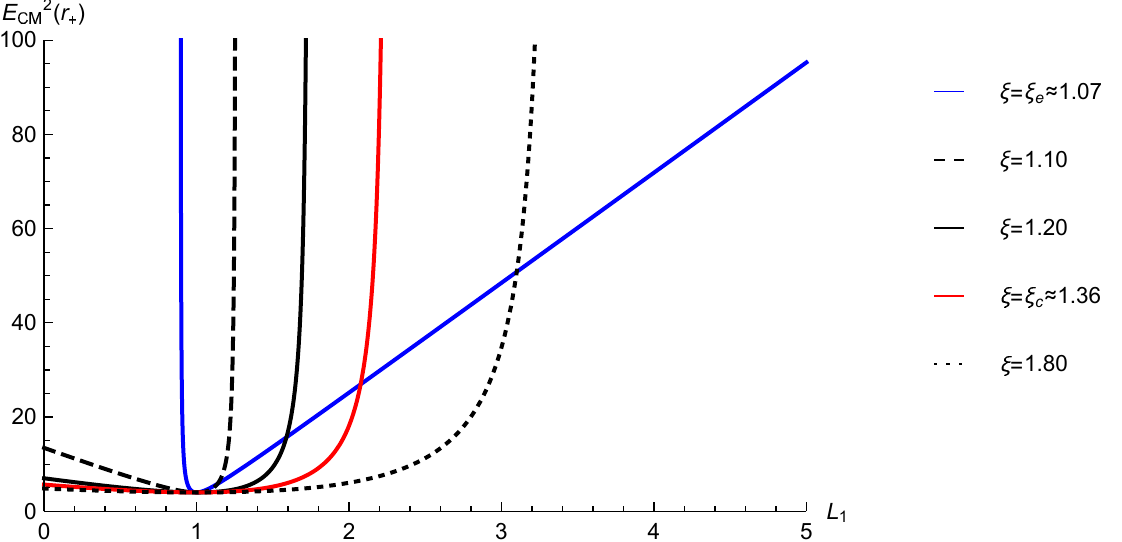}
\end{center}
\caption{The behavior of the CM energy $E_{CM}^2(r_+)$ at the horizon as a function of  $L_1$  for different values of $\xi$ with $M=1$, $\lambda=a=b=1$, $\Lambda=-1$, and $J=1.2$}
\label{f3}
\end{figure}

\begin{figure}[!h]
\begin{center}
\includegraphics[width=100mm]{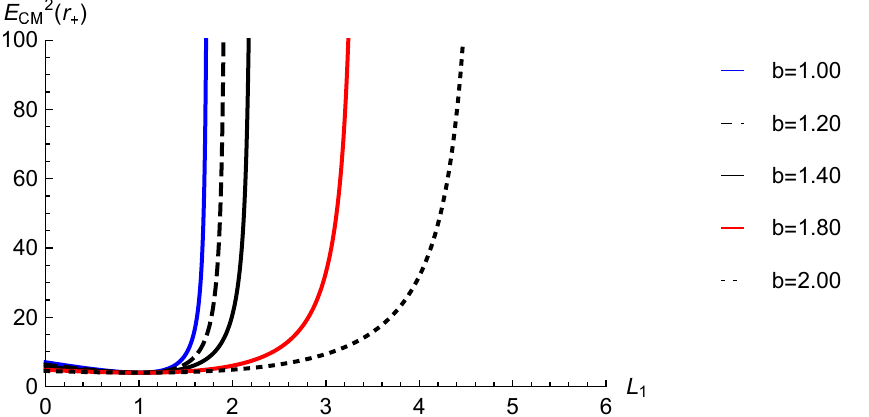}
\includegraphics[width=100mm]{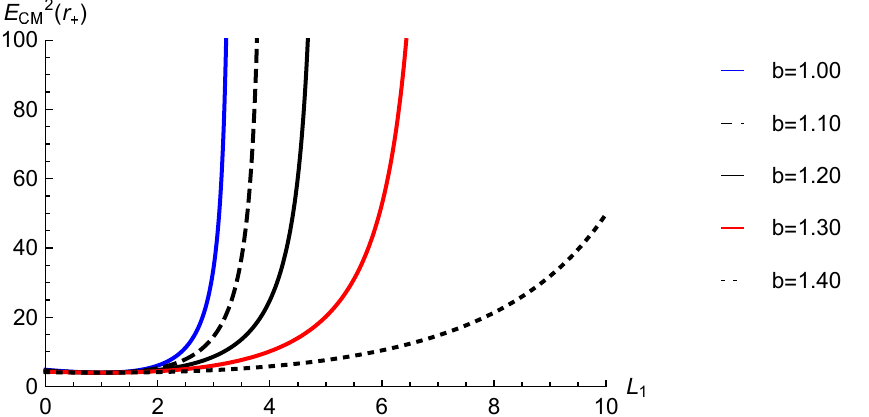}
\end{center}
\caption{The behavior of the CM energy $E_{CM}^2(r_+)$ at the horizon as a function of  $L_1$  for $\xi=1.2$ (top panel) and for $\xi=1.8$ (bottom panel) for different values of $b$ with $M=1$, $\lambda=a=1$, $\Lambda=-1$, and $J=1.2$}
\label{f4}
\end{figure}

\begin{figure}[!h]
\begin{center}
\includegraphics[width=80mm]{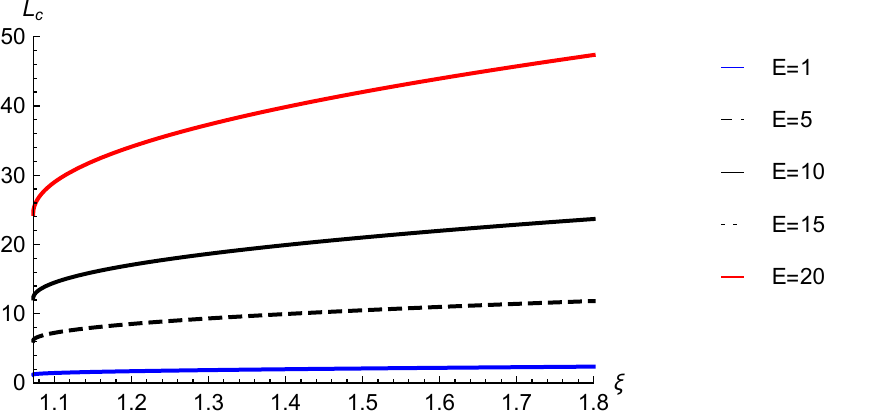}
\end{center}
\caption{The behavior of $L_{c}$ as a function of  $\xi$  for different values of the energy $E=1,5,10,15,20$ with $M=1$, $\lambda=a=b=1$, $\Lambda=-1$, and $J=1.2$}
\label{f5}
\end{figure}

In order to calculate the CM energy of two colliding particles, we use the expressions of the components of the 3-velocity derived in the last section to obtain the CM energy of the colliding particles. Also, we consider that the particles have the same rest mass $m_0$, energies $E_1$ and $E_2$ and angular momentum $L_1$ and $L_2$, respectively. From the relation $E_{CM}=\sqrt{2} m_0 \sqrt{1+g_{\mu \nu}u_1^{\mu}u_2^{\nu}}$, where $u_1$ and $u_2$ denotes the {\bf{3}}-velocities of the particles, we obtain
\begin{equation}\label{CM}
\frac{E_{CM}^2}{2 m_0^{2}}= \frac{4 Z^2 r^2 (r^2 -L_1 L_2 )+(K_1 K_2-H_1 H_2)}{4 Z^2 r^4}~,
\end{equation}
where
\begin{eqnarray}
\notag K_i  &=& 2 E_i r^2-J L_i ~, \\
H_i &=& \sqrt{(2 E_i r^2-J L_i)^2- 4 r^2 Z^2 (r^2+L_i^2)}~,
\end{eqnarray}
and the subscript takes the values $i=1,2$. Also, when the particles arrive to the event horizon $r=r_{+}$, $Z^2(r_+) \rightarrow 0$, $H_1 \rightarrow \sqrt{K_1^2}$ and $H_2 \rightarrow \sqrt{K_2^2}$, the CM energy (\ref{CM}) at the horizon yields:
\begin{equation}
\frac{E_{CM}^2}{2 m_0^{2}}(r \rightarrow r_{+})=\frac{1}{4Z^2 r_{+}^4} (K_1 K_2 -\sqrt{K_1^2} \sqrt{K_2^2})~.
\end{equation}
From the last expression we note that if $K_1K_2<0$ the $E_{CM}^{2}$ on the horizon will be a negative infinity therefore the CM energy will be imaginary and so it is not a physical solution. In fact, we can set $K_1<0$ and $K_2>0$ without loss of generality; however, $K_1<0$ outside the event horizon is in contradiction with condition (\ref{condition1}). However, when $K_1K_2 \geq 0$, the numerator of  this expression will be zero and the value of $E_{CM}$ will be undetermined. 
Now, in order to find the limiting value of the CM energy at the horizon we can use the L'H\^{o}pital rule, obtaining
\begin{equation}
\frac{E_{CM}^2}{2 m_0}=\frac{ r_{+}^2 (K_1(r_+)+K_2(r_+))^2 +(K_1(r_+) L_2-K_2(r_+)L_1)^2}{2 r_{+}^2 K_1(r_+) K_2(r_+)}~.
\end{equation}
Now, the numerator of the above expression is finite at the horizon and if $K_i(r_+)=0$ the CM energy of two colliding particles on the horizon could be arbitrarily high, $E_{CM} \big | _{K_i=0} \rightarrow \infty$. So, from $K_i(r_+)=0$ we obtain that the critical angular momentum is given by:
\begin{equation}\label{critico}
L_{ci}= \frac{2r_+^2E_i}{J}~,  \,\,\,\,\, i=1,2.
\end{equation}
On the other hand, when $K_1(r_+)$ and $K_2(r_+)$ are both zero, then $E_{CM}$ is finite at the horizon. In this case $H_1(r_+)=H_2(r_+)=0$ and
\begin{equation}
\frac{E_{CM}^2}{2 m_0}=1-\frac{L_1 L_2}{r_+^2}~.
\end{equation}
Therefore, in order to obtain an infinite CM energy only one of the colliding particles must have the critical angular momentum, making the BSW process possible.
In Figs. (\ref{f3}) and (\ref{f4}) we show the behavior of $E_{CM}^2(r_+)$ versus $L_{1}$ for different values of the Ho\v{r}ava parameter $\xi$ and aether misalignment parameter $b$, respectively. We observe that there is a critical value of the angular momentum for the particle 1 at which the CM energy blows up. Clearly the figures show up that when the parameter $\xi$ increases, $L_{1}=L_{C}$ increases, the same behavior occurs when the parameter $b$ associated to the aether field  increases. Note that only in the extremal case, the limiting value of the critical angular momentum has to be reached from the left of the asymptotic value in order to get positive CM energy $E_{CM}^2(r_+)$ in all the other cases it has to be reached from the right . Additionally, in Fig. (\ref{f5}), we have plotted $L_{c}$ in terms of the Ho\v{r}ava parameter $\xi$ for different values of the energy $E$. It is shown that the critical angular momentum $L_{c}$ increases when the energy of particle 1 increases or the Ho\v{r}ava parameter increases.

\newpage

\section{Radial motion of the particle with critical angular momentum}
\label{two}
\begin{figure}[!h]
\begin{center}
\includegraphics[width=80mm]{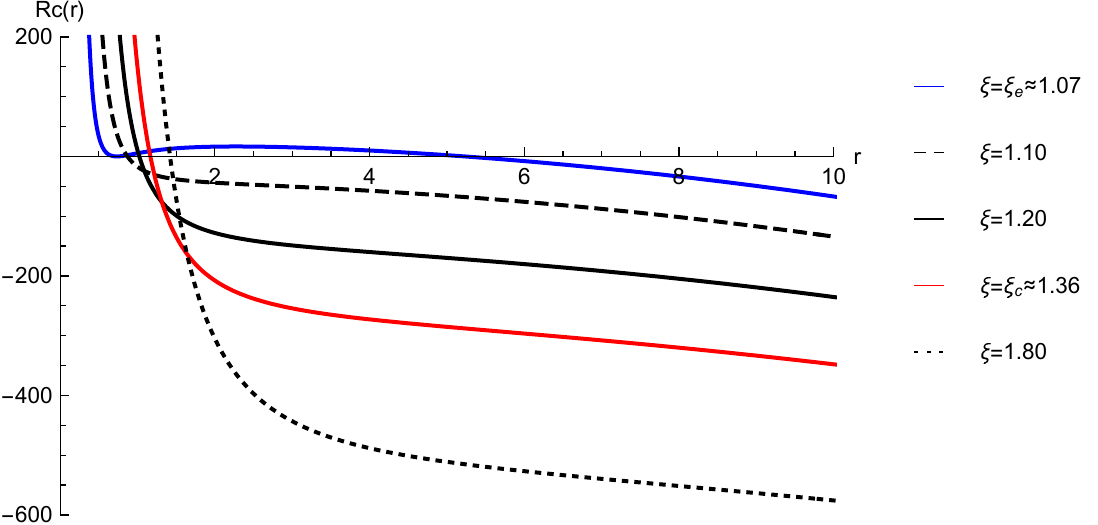}
\includegraphics[width=80mm]{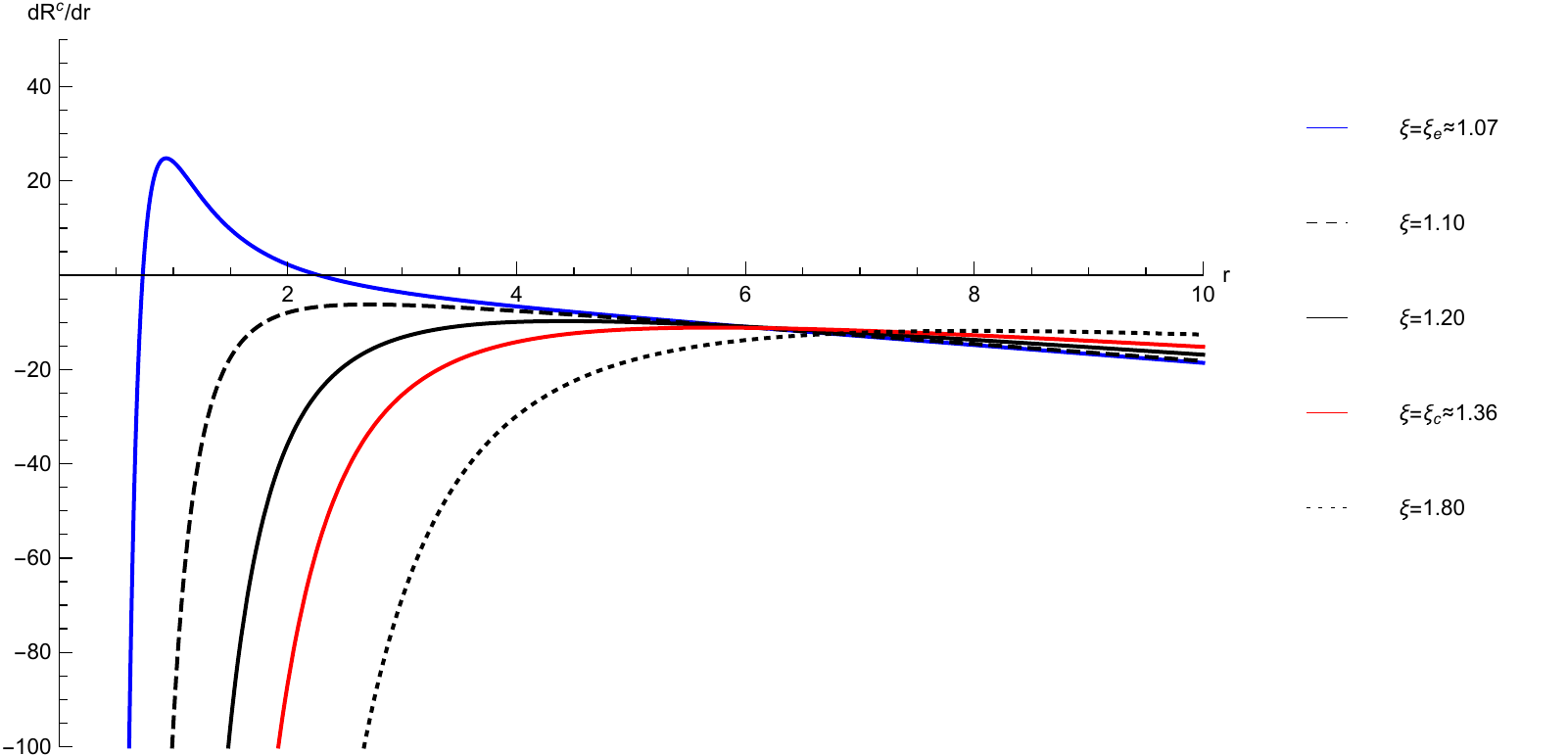}
\end{center}
\caption{The behavior of $R^c(r)$ (left panel) and $dR^c(r)/dr$ (right panel) as a function of $r$ for the extremal (blue line) and non extremal three-dimensional rotating Ho\v{r}ava AdS black hole for different values of the parameter $\xi$, $M=\lambda=a=b=1$, $\Lambda=-1$, $J=1.2$, $m=1$ and $E=10$.}
\label{f6}
\end{figure}

Now, we will study the radial motion of the particle with critical angular momentum and energy $E$. As we have mentioned, the particle can reach the event horizon of the black holes if the square of the radial component of the 3-velocity $\dot{r}^{2}$ in Eq. (\ref{rdot}) is positive or $V$ is negative in the neighborhood of the black hole horizon. We will denote the explicit form of $\dot{r}^{2}$ with critical angular momentum as $R^{c}(r)$, which is given by
\begin{equation}\label{POTC}
R^{c}=\frac{(r^2-r_{+}^2)\left(J^2m^{2}r^{2}(r^2-r_{-}^2)\bar{\Lambda}+E^{2}(J^2(r^2-r_{+}^2)+4(r^2-r_{-}^2)r_{+}^4 \bar{\Lambda})\right)}{J^2 r^4}~,
\end{equation}
and it vanishes on the event horizon. Also, for some values of the parameters, $R^c$ can be positive, which implies that particles with critical angular momentum can exist outside the event horizon; however, as we shall see, they cannot reach the event horizon unless $r_+=r_-$.
Particles with critical angular momentum can reach the event horizon if the following condition
$\frac{dR^{c}}{dr}\Big|_{r=r_{+}}>0$  is satisfied. From (\ref{POTC}) we find
\begin{equation}
\frac{dR^{c}}{dr}\Big|_{r=r_{+}}=\frac{2\bar{\Lambda}(r_{+}^2-r_{-}^2)}{r_{+}}\left(m^2+\frac{4 E^2 r_{+}^2}{J^2}\right)<0~,
\end{equation}
for $r_+ \neq r_-$; therefore, the massive particle with the critical angular
momentum cannot exist just outside of the event horizon $r\gtrsim r_{+}$ if the background is AdS. In Fig. \ref{f6} we plot the behavior of $R^c$ and $\frac{dR^{c}}{dr}$ as a function of $r$ for this background. 

On the other hand, particles with critical angular momentum can reach the Cauchy horizon $r_-$ if $\frac{dR^{c}}{dr}\Big|_{r=r_{-}}>0$. From (\ref{POTC}) we find
\begin{equation}
\frac{dR^{c}}{dr}\Big|_{r=r_{-}}=-\frac{2 (r_+^2-r_-^2)}{r_-} \left( \frac{2 E^2 r_+^2}{r_-^4}+  \bar{\Lambda} \left( 1+ \frac{4 E^2r_+^4}{J^2 r_-^2}  \right) \right)~.
\end{equation}
 Therefore, the massive particle with critical angular
momentum can reach the Cauchy horizon when $-  \bar{\Lambda} \left( 1+ \frac{4 E^2r_+^4}{J^2 r_-^2} \right)>\frac{2 E^2 r_+^2}{r_-^4} $ and the condition that both universal horizon coincide to avoid the zone where aether turns imaginary. For $\bar{J}^2<0$ there is not inner horizon $r_-$, so, the BSW mechanism does not occur.

\begin{figure}[!h]
\begin{center}
\includegraphics[width=80mm]{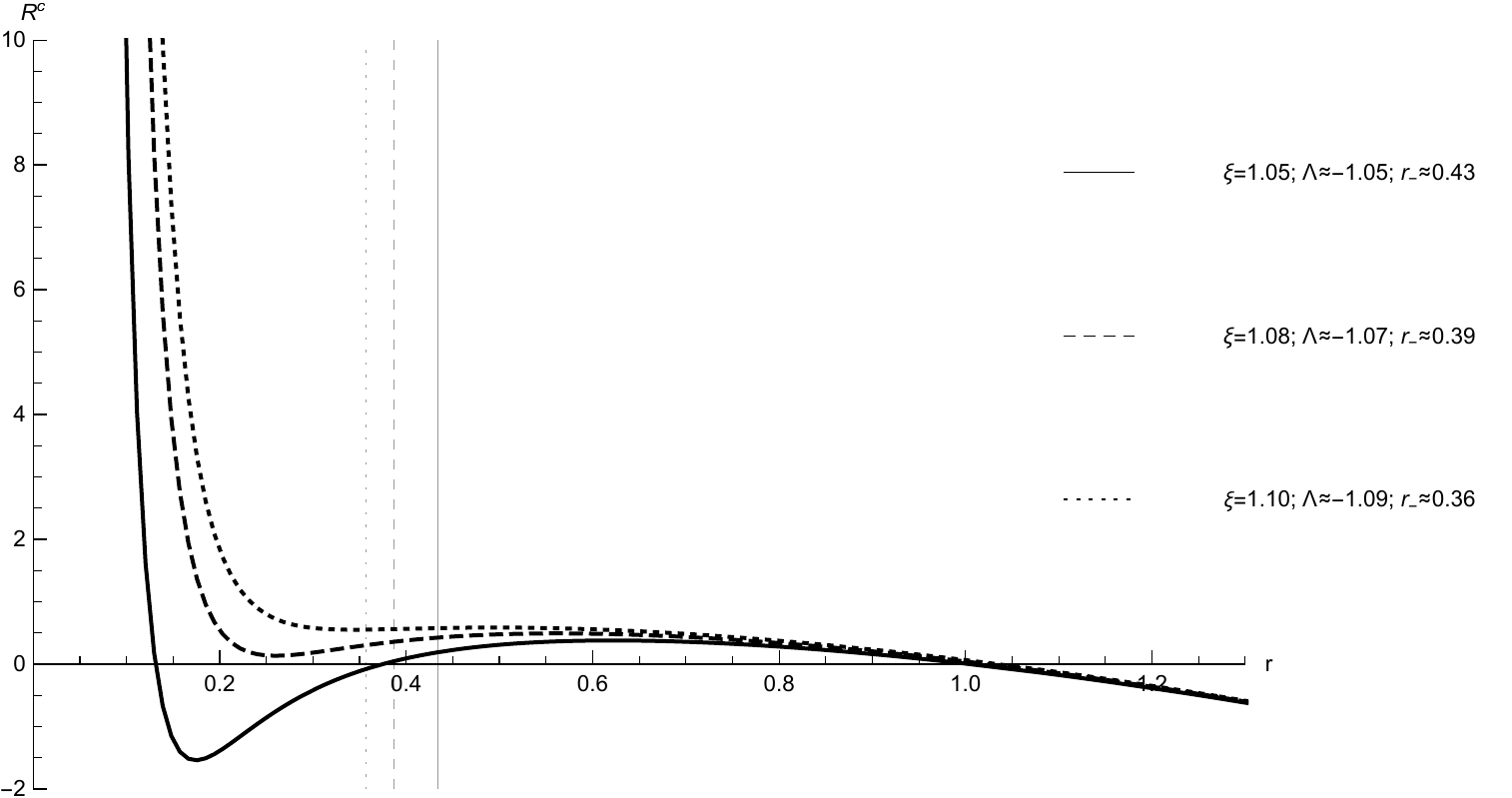}
\includegraphics[width=80mm]{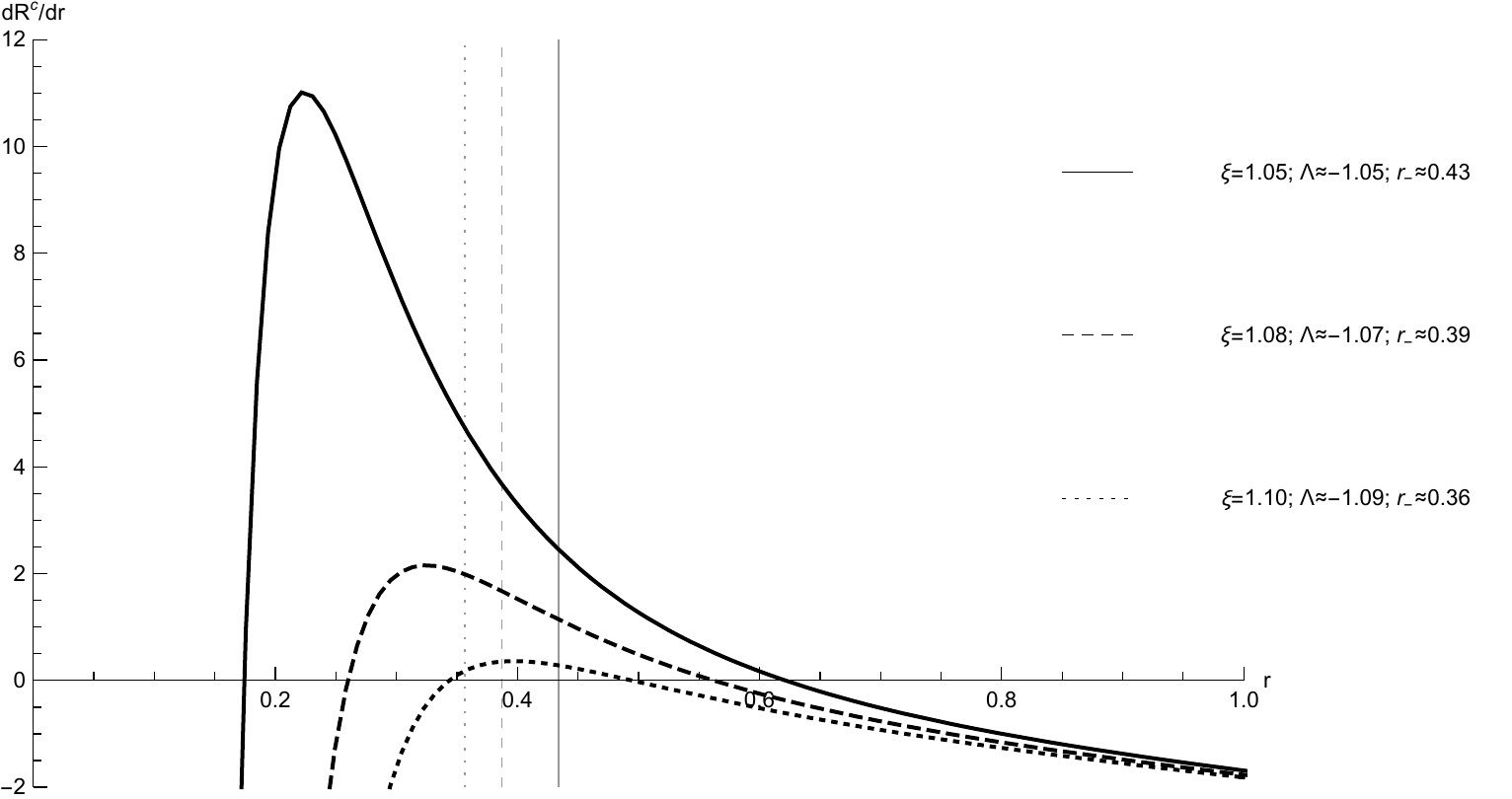}
\end{center}
\caption{The behavior of $R^c(r)$ (left panel) and $dR^c(r)/dr$ (right panel) as a function of $r$ for the non-extremal three-dimensional rotating Ho\v{r}ava AdS black hole for different values of the parameter $\xi$, $M=1.2$, $\lambda=a=b=1$, $\bar{\Lambda}=-1$, $J=1$, $m=1$ and $E=0.10$. In these cases there are not universal horizons, and the vertical lines correspond to the inner horizon, for each black hole solution.}
\label{f7}
\end{figure}

Furthermore, for the extremal black hole $r_{+}=r_{-}$ we obtain
\begin{equation}\label{extremal}
R^{c}=-\frac{(r^2-r_{+}^2)^{2}\left( J^2 m^2 r^2 \bar{\Lambda}+E^2(J^2+4r_{+}^4\bar{\Lambda})\right) }{J^2 r^4}~,
\end{equation}
and
\begin{equation}\label{dextremal}
\frac{dR^{c}}{dr}=-\frac{2(r^2-r_{+}^2)^2\left( J^2 m^2 r^2(r^2+r_{+}^2)\bar{\Lambda}+2 E^2 r_{+}^2(J^2+4r_{+}^4\bar{\Lambda})\right) }{J^2 r^5}~.
\end{equation}
Then, clearly Eqs. (\ref{extremal}) and (\ref{dextremal}) are zero on the event horizon, then it is necessary to calculate $\frac{d^{2}R^{c}}{dr^{2}}\vert_{r=r_{+}}$:
\begin{equation}\label{condition}
\frac{d^{2}R^{c}}{dr^{2}}\Big|_{r=r_{+}}=8\left(m^2 |\bar{\Lambda}|-E^2\frac{(J^2-\bar{J}^2)}{J^2 r_{+}^2}\right) .
\end{equation}
If $\frac{d^{2}R^{c}}{dr^{2}}\Big|_{r=r_{+}}>0$, the particle with critical angular momentum will reach the degenerate horizon. This is fulfilled when
\begin{equation}\label{con1}
E^2< \frac{m^2 |\bar{\Lambda}|J^2 r_{+}^2 }{J^2-\bar{J}^2}\,.
\end{equation}
Therefore, particles with critical angular momentum satisfying the condition (\ref{con1}) will arrive to the degenerate horizon and thus the BSW process is possible. Note also that if $\frac{d^{2}R^{c}}{dr^{2}}>0$, $R^{C}$ has a zero also at $r_{0}=\frac{E \sqrt{J^2-4r_{+}^4|\bar{\Lambda}| }}{Jm\sqrt{|\bar{\Lambda}|}}$ which is grater than $r_{+}$, and the particles with critical angular momentum can exist at $r_{+} \leq r \leq r_{0}$ . 
In Fig. (\ref{f6}) we plot the behavior of $R^c$ and $\frac{dR^c}{dr}$ as a function of $r$ for the  three-dimensional extremal and non-extremal rotating  Ho\v{r}ava AdS black hole.
We observe that 
the particle with critical angular momentum can reach the degenerate horizon if the condition (\ref{con1}) 
is satisfied, and thus the BSW process is possible. It is worth highlighting that this result is different from the usual rotating BTZ black holes ($\xi=1$ and $\lambda=1$), where the massive particles only can exist on the degenerate horizon  \cite{Yang:2012we}. Additionally, Fig. (\ref{f7}) shows up that for the non-extremal case the region where $\frac{dR^c}{dr}>0$ is inside the event horizon $r_+$, in particular $\frac{dR^{c}}{dr}\Big|_{r=r_{-}}>0$, then,  the massive particles with critical angular momentum can reach the inner horizon $r_-$. 

\newpage

\section{Final remarks}
\label{conclusion}

In this paper we considered the collision of two particles in the vicinity of the horizon of a three-dimensional rotating Ho\v{r}ava AdS black hole described by a Lorentz-violating version of the BTZ black hole,  i.e.  a black hole solution with AdS asymptotics, and we analyzed the energy in the CM frame of the colliding particles in order to investigate the effect of the breaking of the Lorentz invariance on the possibility that 
the black hole can act as a particle accelerator. Thus, the differences observed with respect to the BTZ metric are attributed to the breaking of the Lorentz invariance.\\ 

We showed that depending on the parameters, the lapse function can represent a spacetime without an event horizon, i.e. a naked singularity, a black hole geometry with one event horizon, an extremal black hole and finally a black hole with two horizons. Also, one of the most interesting feature within the
context of Lorentz-violating gravity theories is that they
can have universal horizons. Thus, it is possible to observe different zones. One of them, is limited by $r_-$ and $r_+$, and it is described by the existence of the aether, where the roots of $r_u$ are imaginary. Other zones are characterized by two real and distinct universal horizons inside the region between $r_-$ and $r_+$, outside $r_-$, and inside $r_+$; and an especial point where both universal horizons coincide. In the region between $r_u^-$ and $r_u^+$, the aether turns imaginary and the foliation cannot be extended until the singularity. So, if this region is located between the inner and outer horizons then the BSW process is not possible on the inner horizon $r_-$ because the aether is imaginary. We found the following behavior:

\begin{itemize}

    \item
    The existence of a massive particle at infinity depends on the effective cosmological constant ($ \bar{\Lambda}$)  and not on its energy $E$. Thus, massive particles can not exist at infinity for Ho\v{r}ava AdS black holes. 
    \item 
    For a background with positive inner and outer horizons $r_\pm$, $\xi_e < \xi < \xi_c$, we found that the particles with critical angular momentum will never reach the event horizon; therefore, the black hole cannot act as a particle accelerator with unlimited CM energy on the event horizon. On the other hand, the particles can collide on the inner horizon with arbitrarily high CM energy being possible the BSW process. Also, we showed that the critical angular momentum increases when the Ho\v{r}ava parameter $\xi$ increases and the aether parameter $b$ increases. 
    \item
    For the extremal case, $\xi = \xi_e$, we found that the particle with critical angular momentum can exist on the degenerate horizon as long as its conserved energy fulfills the condition $E^2< \frac{m^2 |\bar{\Lambda}|J^2 r_{+}^2 }{J^2-\bar{J}^2}$  with the BSW process being possible. Also, we showed that the critical particle can exist between  $r_{+} \leq r \leq r_{0}$, with $r_{0}=\frac{E \sqrt{J^2-4r_{+}^4|\bar{\Lambda}| }}{Jm\sqrt{|\bar{\Lambda}|}}$, this result is different to the extremal BTZ black holes where particles with critical angular momentum only can exist on the degenerate horizon. 
    \item
    For $\xi=\xi_c$, that is, $r_-=0$, the two sectors converge. This occurs when  $J^2+4a^2(1-\xi)=0$ $(\bar{J}=0)$. Also, the fine tuning of the critical angular momentum $L_{c}$ is bigger than the extremal and the non-extremal cases, nevertheless the particle with $L_{c}$ will never reach the event horizon.
    \item
    In the range $\xi > \xi_c$ the black hole only has one horizon $r_{+}>0$.  For  this case  $J^2<4a^2(\xi-1)$ and the BSW mechanism is not possible.
\end{itemize}

\section*{Acknowledgments}

Y.V. acknowledge support by the Direcci\'on de Investigaci\'on y Desarrollo de la Universidad de La Serena, Grant No. PR18142. 
P.A.G. would like to thank the Facultad de Ciencias, Universidad de La Serena for its hospitality. Y.V. and R.B. would like to thank the Facultad de Ingenier\'{i}a y Ciencias, Universidad Diego Portales for its hospitality.


\begin{thebibliography}{99}


\bibitem{Banados:2009pr} 
  M.~Ba\~nados, J.~Silk and S.~M.~West,
  Phys.\ Rev.\ Lett.\  {\bf 103}, 111102 (2009)
  [arXiv:0909.0169 [hep-ph]].
 
  \bibitem{PS1} 
T. Piran, J. Shaham and J. Katz. 1975. Astrophys.J.,196,L107

\bibitem{Piran:1977dm} 
  T.~Piran and J.~Shaham,
  Phys.\ Rev.\ D {\bf 16}, 1615 (1977).
  doi:10.1103/PhysRevD.16.1615
  
  \bibitem{PS3} 
T. Piran and J. Shaham. 1977. Astrophys.J.,214,268


\bibitem{Grib:2010at} 
  A.~A.~Grib and Y.~V.~Pavlov,
Int.\ J.\ Mod.\ Phys.\ D {\bf 20}, 675 (2011)
[arXiv:1008.3657 [gr-qc]].
  
 
 
\bibitem{Zaslavskii:2010jd} 
  O.~B.~Zaslavskii,
Phys.\ Rev.\ D {\bf 82}, 083004 (2010)
[arXiv:1007.3678 [gr-qc]].
 
\bibitem{Zaslavskii:2010aw} 
  O.~B.~Zaslavskii,
JETP Lett.\  {\bf 92}, 571 (2010)
[Pisma Zh.\ Eksp.\ Teor.\ Fiz.\  {\bf 92}, 635 (2010)]
[arXiv:1007.4598 [gr-qc]].
  
  
  
\bibitem{Gao:2011sv} 
  S.~Gao and C.~Zhong,
Phys.\ Rev.\ D {\bf 84}, 044006 (2011)
[arXiv:1106.2852 [gr-qc]].
 
\bibitem{Li:2010ej} 
  Y.~Li, J.~Yang, Y.~L.~Li, S.~W.~Wei and Y.~X.~Liu,
Class.\ Quant.\ Grav.\  {\bf 28}, 225006 (2011)
[arXiv:1012.0748 [hep-th]].



\bibitem{Said:2011qm} 
  J.~L.~Said and K.~Z.~Adami,
Phys.\ Rev.\ D {\bf 83}, 104047 (2011)
[arXiv:1105.2658 [gr-qc]].
  

\bibitem{Zhong:2011vq} 
  C.~Zhong and S.~Gao,
JETP Lett.\  {\bf 94}, 589 (2011)
[arXiv:1109.0772 [hep-th]].

\bibitem{Yi:2011}
Y.~Zhu, S.~Fengwu,Y.~Xiao Liu,Y.~Jiang,
Phys.Rev.D84:043006,2011 

  
\bibitem{Abdujabbarov:2013qka} 
  A.~Abdujabbarov, N.~Dadhich, B.~Ahmedov and H.~Eshkuvatov,
  Phys.\ Rev.\ D {\bf 88}, 084036 (2013)
  [arXiv:1310.4494 [gr-qc]].
  
 
  
  \bibitem{Galajinsky:2013as} 
  A.~Galajinsky,
  Phys.\ Rev.\ D {\bf 88}, 027505 (2013)
  [arXiv:1301.1159 [gr-qc]].

\bibitem{Shaymatov:2013tna} 
  S.~R.~Shaymatov, B.~J.~Ahmedov and A.~A.~Abdujabbarov,
  Phys.\ Rev.\ D {\bf 88}, no. 2, 024016 (2013).
  doi:10.1103/PhysRevD.88.024016


\bibitem{Sadeghi:2013gmf} 
  J.~Sadeghi, B.~Pourhassan and H.~Farahani,
  Commun.\ Theor.\ Phys.\  {\bf 62}, no. 3, 358 (2014)
  [arXiv:1310.7142 [hep-th]].
  
\bibitem{Fernando:2013qba} 
  S.~Fernando,
  Gen.\ Rel.\ Grav.\  {\bf 46}, 1634 (2014)
  [arXiv:1311.1455 [gr-qc]].





\bibitem{Ghosh:2014mea} 
  S.~G.~Ghosh, P.~Sheoran and M.~Amir,
  Phys.\ Rev.\ D {\bf 90}, no. 10, 103006 (2014)
  [arXiv:1410.5588 [gr-qc]].
  
  
  
 
\bibitem{Pradhan:2014eza} 
  P.~Pradhan,
Astropart.\ Phys.\  {\bf 62}, 217 (2015)
[arXiv:1407.0877 [gr-qc]].
  
\bibitem{Ghosh:2015pra} 
  S.~G.~Ghosh and M.~Amir,
  Eur.\ Phys.\ J.\ C {\bf 75}, no. 11, 553 (2015)
  [arXiv:1506.04382 [gr-qc]].
  
  
\bibitem{Amir:2015pja} 
  M.~Amir and S.~G.~Ghosh,
  JHEP {\bf 1507}, 015 (2015)
  [arXiv:1503.08553 [gr-qc]].
  
  
  
\bibitem{Guo:2016vbt} 
  M.~Guo and S.~Gao,
Phys.\ Rev.\ D {\bf 93}, no. 8, 084025 (2016)
[arXiv:1602.08679 [gr-qc]].


 
\bibitem{Zhang:2016btg} 
  Y.~P.~Zhang, B.~M.~Gu, S.~W.~Wei, J.~Yang and Y.~X.~Liu,
  Phys.\ Rev.\ D {\bf 94}, no. 12, 124017 (2016)
  [arXiv:1608.08705 [gr-qc]].
  
  
  
  
   
\bibitem{Armaza:2015eha} 
  C.~Armaza, M.~Ba\~nados and B.~Koch,
Class.\ Quant.\ Grav.\  {\bf 33}, no. 10, 105014 (2016)
[arXiv:1510.01223 [gr-qc]].

 
\bibitem{Zaslavskii:2016dfh} 
  O.~B.~Zaslavskii,
EPL {\bf 114}, no. 3, 30003 (2016)
[arXiv:1603.09353 [gr-qc]].
  
  
\bibitem{Zaslavskii:2016stw} 
  O.~B.~Zaslavskii,
  Int.\ J.\ Mod.\ Phys.\ D {\bf 26}, no. 10, 1750108 (2017)
  [arXiv:1602.08779 [gr-qc]].
  
  

\bibitem{Fernando:2017kut} 
  S.~Fernando,
  Mod.\ Phys.\ Lett.\ A {\bf 32}, 1750074 (2017)
  [arXiv:1703.00373 [gr-qc]].

\bibitem{Gonzalez:2018lfs} 
  P.~A.~González, M.~Olivares, E.~Papantonopoulos and Y.~Vásquez,
  Phys.\ Rev.\ D {\bf 97}, no. 6, 064034 (2018)
  [arXiv:1802.01760 [gr-qc]].

\bibitem{Gonzalez:2018zuu} 
  P.~A.~González, M.~Olivares, Y.~Vásquez, J.~Saavedra and A.~Övgün,
  Eur.\ Phys.\ J.\ C {\bf 79}, no. 6, 528 (2019)
  [arXiv:1811.08551 [gr-qc]].
  
  
\bibitem{Jiang:2019cuc} 
  J.~Jiang and S.~Gao,
  Eur.\ Phys.\ J.\ C {\bf 79}, no. 5, 378 (2019)
  [arXiv:1905.02491 [hep-th]].
 




 

  
  


\bibitem{ata}  
 F. Atamurotov, B. Ahmedov and S. Shaymatov,
 Astrophys. Space. Sci. {\bf 347}, 277 (2013)
 
\bibitem{Horava:2009uw} 
  P.~Horava,
  Phys.\ Rev.\ D {\bf 79}, 084008 (2009)
  [arXiv:0901.3775 [hep-th]].
  
  
 
\bibitem{Sotiriou:2011dr}
  T.~P.~Sotiriou, M.~Visser and S.~Weinfurtner,
 ``Lower-dimensional Horava-Lifshitz gravity,''
  Phys.\ Rev.\ D {\bf 83}, 124021 (2011)
  [arXiv:1103.3013 [hep-th]].
  
\bibitem{Sotiriou:2014gna} 
  T.~P.~Sotiriou, I.~Vega and D.~Vernieri,
  ``Rotating black holes in three-dimensional Hořava gravity,''
  Phys.\ Rev.\ D {\bf 90}, no. 4, 044046 (2014)
  [arXiv:1405.3715 [gr-qc]].


 

 \bibitem{Becar:2019hwk} 
  R.~Bécar, P.~A.~González, E.~Papantonopoulos and Y.~Vásquez,
  arXiv:1906.06654 [gr-qc].

\bibitem{Gonzalez:2019xfr} 
  P.~A.~González, M.~Olivares, E.~Papantonopoulos and Y.~Vásquez,
  arXiv:1912.00946 [gr-qc].
  
  


  \bibitem{Zhu:2019ura} 
  T.~Zhu, Q.~Wu, M.~Jamil and K.~Jusufi,
  ``Shadows and deflection angle of charged and slowly rotating black holes in Einstein-Æther theory,''
  Phys.\ Rev.\ D {\bf 100}, no. 4, 044055 (2019)



\bibitem{Abdujabbarov:2011af} 
  A.~Abdujabbarov, B.~Ahmedov and B.~Ahmedov,
  Phys.\ Rev.\ D {\bf 84}, 044044 (2011)
  [arXiv:1107.5389 [astro-ph.SR]].
  
  
  

  

\bibitem{Sadeghi:2011qu} 
  J.~Sadeghi and B.~Pourhassan,
  Eur.\ Phys.\ J.\ C {\bf 72}, 1984 (2012)
  [arXiv:1108.4530 [hep-th]].
  
\bibitem{Halilsoy:2015rna} 
  M.~Halilsoy and A.~Ovgun,
  Adv.\ High Energy Phys.\  {\bf 2017}, 4383617 (2017)
  [arXiv:1504.03840 [gr-qc]].
  
  
  
\bibitem{Lake:2010bq} 
  K.~Lake,
  Phys.\ Rev.\ Lett.\  {\bf 104}, 211102 (2010)
  Erratum: [Phys.\ Rev.\ Lett.\  {\bf 104}, 259903 (2010)]
  [arXiv:1001.5463 [gr-qc]].
  
  
  
  
  
\bibitem{Yang:2012we} 
  J.~Yang, Y.~L.~Li, Y.~Li, S.~W.~Wei and Y.~X.~Liu,
  Adv.\ High Energy Phys.\  {\bf 2014}, 204016 (2014)
  [arXiv:1202.4159 [hep-th]].

\bibitem{Tsukamoto:2017rrl} 
  N.~Tsukamoto, K.~Ogasawara and Y.~Gong,
  Phys.\ Rev.\ D {\bf 96}, no. 2, 024042 (2017)
  [arXiv:1705.10477 [gr-qc]].
  
  
  

\bibitem{Becar:2017aag} 
  R.~Bécar, P.~A.~González and Y.~Vásquez,
  Eur.\ Phys.\ J.\ C {\bf 78}, no. 4, 335 (2018)
  [arXiv:1712.00868 [gr-qc]].
  
  
  
 
  
  

\bibitem{Jacobson:2010mx} 
  T.~Jacobson,
  Phys.\ Rev.\ D {\bf 81}, 101502 (2010)
  Erratum: [Phys.\ Rev.\ D {\bf 82}, 129901 (2010)]
  [arXiv:1001.4823 [hep-th]].
  
  
  
  \bibitem{Cropp:2013sea} 
 
  B.~Cropp, S.~Liberati, A.~Mohd and M.~Visser,
  Phys.\ Rev.\ D {\bf 89}, no. 6, 064061 (2014)
  [arXiv:1312.0405 [gr-qc]].
 
  \bibitem{chandra}
Chandrasekhar S.:
The Mathematical Theory of Black Holes.
Oxford University Press, New York (1983).


\end{thebibliography}
\end{document}